\documentclass[a4paper,12pt]{article}

\usepackage[utf8]{inputenc}
\usepackage{graphicx} % inclusão de figuras
\usepackage[T1]{fontenc} 
\usepackage{amsmath}
\usepackage{amssymb}
\usepackage{hyperref}
\usepackage{lmodern}
\usepackage{empheq}
\usepackage{braket}
\usepackage{xcolor}
\usepackage{cancel}
\usepackage{cite}
\usepackage{scalerel}
\usepackage{afterpage}
\usepackage{cleveref}
\usepackage[mathscr]{eucal}
\usepackage[shortlabels]{enumitem}
\usepackage{multirow}
\usepackage[left=2.5cm,right=2.5cm,top=2.5cm,bottom=2.5cm]{geometry}

\definecolor{royalblue}{rgb}{0.0, 0.14, 0.4}

\hypersetup{
	colorlinks = true,
	allcolors = {blue},
}

\begin{document}

\begin{titlepage}
\vspace{2.5cm}	
\begin{center}
\Large \textbf{Asymptotic Stability Analysis for $SU(n)$ Dark Monopoles}
\end{center}
\vspace{0.3cm}
\begin{center} \normalsize
M.L. Deglmann\footnote{m.l.deglmann@posgrad.ufsc.br} and Marco A.C. Kneipp\footnote{marco.kneipp@ufsc.br}
\end{center}
\setcounter{footnote}{0}
\vspace{0.2cm}
\begin{center}
Universidade Federal de Santa Catarina (UFSC),\\
Departamento de Física,\\
Campus Universitário, Trindade,\\
88040-900, Florianópolis, Brazil.
\end{center}
\vspace{0.3cm}
\begin{abstract}
We analyze the asymptotic stability of the $SU(n)$ Dark Monopole solutions and we show that there are unstable modes associated with them. We obtain the explicit form of the unstable perturbations and the associated negative-squared eigenfrequencies.
\end{abstract}
\vspace{\fill}
\texttt{Keywords:} Magnetic Monopoles, Gauge Theories, Stability.
\vspace{0.5cm}
\end{titlepage}

\section{Introduction}

It is believed that the Standard Model can be embedded in a Grand Unified Theory (GUT). In many of these GUTs, there is a scalar field $\phi$ in the adjoint representation which produces a symmetry breaking of the gauge group $G$ of the form\footnote{Quotation marks refer to the local structure of the unbroken gauge group, only.} $G\rightarrow G_{0}=``K\times U(1)"$, with a compact $U(1)$, allowing the existence of topological monopoles \cite{CorriganOlive76,GO78,GO81}. In general, these monopole solutions possess a magnetic field taking values in the abelian subalgebra $h_{0}$ of the unbroken gauge group $G_{0}$. These solutions are generalizations of the 't Hooft-Polyakov monopole \cite{'tHooft,Polyakov}. On the other hand, in \cite{DeglmannKneipp} we have found monopole solutions whose magnetic field is not in the Cartan subalgebra $h_{0}$. As a consequence, these monopoles do not possess a magnetic flux in the direction of the electromagnetic $U(1)_{em}$ generator. Thus, we have named them Dark Monopoles. These monopole solutions were constructed for the case of a Yang-Mills-Higgs theory with an arbitrary gauge group $G=SU(n)$ broken to
\begin{equation*}
G_{0}=[U(1)\times SU(p)\times SU(n-p)]/Z\,,
\end{equation*} 
by a vacuum configuration $\phi_{0}=v\,\lambda_{p}\cdot H/|\lambda_{p}|$, where $\lambda_{p}$ is a fundamental weight of $SU(n)$ and $v$ is the vacuum expectation value (VEV). Therefore, as a particular case, these Dark Monopoles can exist in $SU(5)$ GUT.

Before we describe the summary of our construction, let us recall that for an arbitrary root $\alpha$ we  can define the generators\footnote{In this paper, we shall adopt conventions and definitions of reference \cite{DeglmannKneipp}.} 
\begin{equation}
\begin{aligned}
T_{1}^{\alpha} &
=\frac{E_{\alpha}+E_{-\alpha}}{2}\,,\qquad
T_{2}^{\alpha} & =\frac{E_{\alpha}-E_{-\alpha}}{2i}\,,\qquad
T_{3}^{\alpha} &
=\frac{\alpha\cdot H}{\alpha^{2}}\,,
\end{aligned}\label{su2_subalgebras}
\end{equation}
which form an $su(2)$ subalgebra. In what follows, we shall work with the roots of $su(n)$, which can be written as $\alpha_{ij} = e_{i}-e_{j}$, $i,j=1,2,\dots, n$, where $\left(e_{i}\right)_{k}=\delta_{ik}$ are the unit vectors in the $n$-dimensional vector space. Hence, we can use the indices of these roots to name the generators, e.g., $T_{a}^{ij}$ ($a=1,2,3$) are the generators of \cref{su2_subalgebras} associated with the root $\alpha_{ij}$.

We have constructed the Dark Monopole solutions associated with an $SO(3)$ subgroup, whose generators are 
\begin{align}
M_{3} & =2T_{2}^{ij}\,,\notag \\
M_{1} & =2T_{2}^{jk}\,,\label{eq:su(2) positive alpha}\\
M_{2} & =2T_{2}^{ki}\,,\notag 
\end{align}
which we have called monopole generators. Note that $M_{3}$ belongs to the unbroken group $G_{0}$, which implies that $\alpha_{ij}\cdot\lambda_{p}=0$, while $M_{1}$ and $M_{2}$ are broken generators. 

Now, we recall that for $r>R_{c}$, where $R_{c}$ is the monopole core, our solutions have the following asymptotic configuration
\begin{equation}
\begin{aligned}
W_{i} & =-\frac{1}{er}\epsilon_{ijk}n^{j}M_{k},\label{ansatz-DM}\\
\phi & =v\,S+\alpha\,\sum_{m=-2}^{2}Y_{2,m}^{*}(\theta,\varphi)\,Q_{m}\,,
\end{aligned}
\end{equation}
where $\alpha=\frac{2v}{\sqrt{6}\,|\lambda_{p}|}\,\sqrt{\frac{4\pi}{5}}$ and $e$ is the coupling constant of the theory. The generators $Q_{m}$ form a quintuplet under the $su(2)$ algebra of \cref{eq:su(2) positive alpha} and are given by 
\begin{align*}
Q_{0} & =\frac{2}{\sqrt{6}}\left(T_{3}^{ik}+T_{3}^{jk}\right)\,,\\
Q_{\pm1} & =\pm\left(T_{1}^{ik}\pm i\,T_{1}^{jk}\right)\,,\\
Q_{\pm2} & =-\left(T_{3}^{ij}\pm i\,T_{1}^{ij}\right)\,.
\end{align*}
Besides that, $S$ is a singlet. The gauge field configuration gives rise to the asymptotic magnetic field 
\begin{equation*}
B_{i}=-\frac{n^{i}}{er^{2}}n^{a}M_{a}=-\frac{x^{i}}{er^{3}}gM_{3}g^{-1}\,,
\end{equation*}
where $n^{i}=x^{i}/r$ and 
\begin{equation}
g=\exp\left(-i\varphi M_{3}\right)\exp\left(-i\theta M_{2}\right)\exp\left(i\varphi M_{3}\right)\,.\label{g-theta-phi}
\end{equation}
It is relevant to note that, for the Dark Monopoles, the asymptotic magnetic field has the traditional hedgehog form in the subspace generated by the $M_{i}$ generators, while the scalar field belongs to the $5$-dimensional subspace generated by the $Q_{m}$ generators.

Another relevant point to be resumed is that the Dark Monopoles possess an asymptotic symmetry associated with a Killing vector $\zeta$, so that $\zeta=n^{a}M_{a}$ outside the monopole core. In addition, there is a gauge-invariant conserved current given by
\begin{equation*}
J_{M}^{\mu}\equiv\frac{1}{|\overline{\zeta}|y}\,\partial_{\nu}\text{Tr}\left(^{*}G^{\mu\nu}\zeta\right)\,,\label{4-current}
\end{equation*}
where $|\bar{\zeta}|=1$ and $\psi^{2}y$ is the Dynkin index of the representation, while $\psi$ is the highest root of the Lie algebra of $G$, $L(G)$. The current $J_{M}^{\mu}$ implies a non-abelian magnetic charge $Q_{M} = -8\pi/e$. 
On the other hand, we have also shown that $Q_{M}$ can be understood in a geometrical way, in terms of the Poincaré-Hopf index for $\zeta$. In fact, this topological quantity measures the number of times $\hat{\zeta}$ covers a $2$-sphere in internal space as $\hat{r}$ covers $S^{2}_{\scaleto{\infty}{3.5pt}}$ once. Again, we recall that the magnetic charge of these Dark Monopoles is not the usual one in the abelian direction, but a non-abelian case in the $\zeta-$direction.

Nevertheless, the question regarding the stability of these Dark Monopoles was still open. In the present paper, we shall analyze this problem. In order to do that, we shall proceed with an asymptotic stability analysis to understand whether the conservation of this non-abelian magnetic charge can make the Dark Monopoles stable. The problem of stability for monopole solutions was first investigated by Brandt and Neri \cite{BrandtNeri} and Coleman \cite{Coleman50}, for the case of singular non-abelian monopoles. The case of non-singular solutions was analyzed by \cite{Zhang,Horvathy88}, while the specific case of the 't Hooft-Polyakov monopole was investigated by \cite{Baacke}. Here we follow a different approach to this problem, even though we recover the results of \cite{BrandtNeri,Coleman50,Shnir}. Finally, we note that stability analysis was also applied to other topological solutions \cite{EarnshawJames,JamesPeriVachaspati,Hollmann,Hartmann,BazeiaSantos,AchucarroUrrestilla}.

This paper is organized as follows: in \cref{Linear-Perturbations}, we obtain the equations for the perturbations of the Higgs and gauge fields. In \cref{Radial-Equations-Section}, we solve the equation for the gauge field perturbation and obtain a general result for all the unstable eigenfrequencies which may occur for $SU(n)$ Dark Monopoles. In \cref{Unstable-Modes-Section}, we obtain the explicit form of the unstable perturbations. Finally, in \cref{Discussions-and-Conclusion-Section} we present our discussions and conclusions. 

\section{Linear Perturbations}\label{Linear-Perturbations}

In this section, we shall analyze the equations for the perturbations of the fields of $SU(n)$ Dark Monopoles, which is a generalization of the 't Hooft-Polyakov case (see, for instance, \cite{Shnir} and references therein).

Given the Dark Monopole solution, let us consider the perturbed configuration
\begin{align}
\phi &=\bar{\phi}+\delta\phi\,,\\
W_{i} &=\bar{W}_{i}+\delta W_{i}\,,
\end{align}
where $\bar{\phi},\,\bar{W}_{i}$ are the original fields of the Dark Monopole solution, given by \cref{ansatz-DM}. The perturbations are such that $\delta\phi=\delta\phi(x,t)$ and $\delta W_{i}=\delta W_{i}(x,t)$. Moreover, 
\begin{equation*}
D_{i}=\bar{D}_{i}+ie\,[\delta W_{i},\,\cdot\,\,]\,,
\end{equation*}
with $\bar{D}_{i}=\partial_{i}+ie\,[\bar{W}_{i},\,\cdot\,\,]$.
Hence, up to first order in perturbations, 
\begin{align*}
D_{i}\phi &=\bar{D}_{i}\bar{\phi}+\bar{D}_{i}\delta\phi+ie\,[\delta W_{i},\bar{\phi}]\,,\\%\label{D_i_phi}\\
G_{ij}&=
\bar{G}_{ij}+\bar{D}_{i}\delta W_{j}-\bar{D}_{j}\delta W_{i}\,,%\label{G_ij}
\end{align*}
implying that
\begin{align}
D_{i}G_{ij}
&=\bar{D}_{i}\bar{G}_{ij}+\bar{D}_{i}\bar{D}_{i}\delta W_{j}-\bar{D}_{i}\bar{D}_{j}\delta W_{i}-ie[\bar{G}_{ij},\delta W_{i}]\,,\label{D_i_G_ij}\\
ie\,[\phi,D_{j}\phi]
&=ie\,[\bar{\phi},\bar{D}_{j}\bar{\phi}]-\bar{D}_{j}\bar{D}_{i}\delta W_{i} - 2ie\,[\bar{D}_{j}\bar{\phi},\delta\phi] + e^{2}\,[\bar{\phi},[\bar{\phi},\delta W_{j}]]\,.\label{comut-phi-D_j_phi}
\end{align}
Note that, in order to obtain eq.\eqref{comut-phi-D_j_phi}, we have used the background gauge condition \cite{WeinbergYi}
\begin{equation}
\bar{D}_{i}\delta W_{i}+ie\,[\bar{\phi},\delta\phi]=0\,,\label{background-gauge-condition}
\end{equation}
so that we eliminate redundancies associated with the gauge freedom. The surviving gauge modes correspond to global gauge transformations in the unbroken subgroup, which are physically relevant.

Plugging these results into the equation of motion 
\begin{equation*}
D_{\mu}G^{\mu\nu}=-ie\,[\phi,D^{\nu}\phi]\,,
\end{equation*}
while using the fact that $W_{0}=0=\delta W_{0}$, we obtain that 
\begin{equation}
\partial_{0}^{2}\,\delta W_{j}=\bar{D}_{i}\bar{D}_{i}\delta W_{j}+2ie\,\epsilon_{kij}\,[\bar{B}_{k},\delta W_{i}]+2ie\,[\bar{D}_{j}\bar{\phi},\delta\phi]-e^{2}\,[\bar{\phi},[\bar{\phi},\delta W_{j}]]\,.\label{eq-pert-W-full}
\end{equation}
We shall only consider perturbations to the asymptotic monopole configuration, i.e. outside the monopole core, where $r>R_{c}$. In this region, $\bar{B}_{k}=-\frac{n^{k}}{er^{2}}\,n^{l}M_{l}$
and $\bar{D}_{j}\bar{\phi}=0$, which implies that the above equation simplifies to 
\begin{equation}
\partial_{0}^{2}\,\delta W_{i}=\bar{D}_{m}\bar{D}_{m}\delta W_{i} - 2i\,\epsilon_{ijk}\,\frac{n^{j}}{r^{2}}\,[n^{l}M_{l},\delta W_{k}] - e^{2}\,[\bar{\phi},[\bar{\phi},\delta W_{i}]]\,.\label{eq-pert-W}
\end{equation}

On the other hand, from the equation of motion for the Higgs field\footnote{By $\phi^{2}$ we mean $\text{Tr}\left(\frac{\phi\phi}{y}\right)$.}
\begin{equation*}
D_{\mu}\left(D^{\mu}\phi\right)=-2\lambda\left(\phi^{2}-v^{2}\right)\phi\,,
\end{equation*}
we can see that the scalar field perturbation satisfies
\begin{equation*}
\partial_{0}^{2}\delta\phi-D_{i}\left(D_{i}\phi\right)=-2\lambda\left(\phi^{2}-v^{2}\right)\phi\,.\label{pert-phi}
\end{equation*}
Then, it follows that 
\begin{equation*}
\partial_{0}^{2}\,\delta\phi = \bar{D}_{i}\bar{D}_{i}\delta\phi - 2ie\,[\bar{D}_{i}\bar{\phi},\delta W_{i}] - e^{2}\,[\bar{\phi},[\bar{\phi},\delta\phi]] - 2\lambda\left(\frac{1}{y}\text{Tr}\left(\bar{\phi}\bar{\phi}\right)-v^{2}\right)\delta\phi-\frac{4\lambda}{y}\,\text{Tr}\left(\bar{\phi}\delta\phi\right)\bar{\phi}\,.
\end{equation*}
Once more, for the asymptotic configuration, the above equation simplifies to 
\begin{equation}
\partial_{0}^{2}\delta\phi=\bar{D}_{i}\bar{D}_{i}\delta\phi-e^{2}\,[\bar{\phi},[\bar{\phi},\delta\phi]]-\frac{4\lambda}{y}\,\text{Tr}\left(\bar{\phi}\delta\phi\right)\bar{\phi}\,.\label{eq-pert-phi}
\end{equation}
Now, the structure of \cref{eq-pert-W,eq-pert-phi} suggests that the spatial and temporal dependence can be factorized as  
\begin{align*}
\delta W_{i}(x,t) 
&=\exp\left(i\omega_{{\scaleto{G}{3.5pt}}}t\right)\delta W_{i}\left(x\right)\,,\\
\delta\phi(x,t) 
&=\exp\left(i\omega_{\phi}t\right)\delta\phi(x)\,,
\end{align*}
which implies that they can be written as
\begin{align}
-\omega_{{\scaleto{G}{3.5pt}}}^{2}\delta W_{i} 
&=\bar{D}_{m}\bar{D}_{m}\delta W_{i} - 2i\,\epsilon_{ijk}\,\frac{n^{j}}{r^{2}}\,[n^{l}M_{l},\delta W_{k}] - e^{2}\,[\bar{\phi},[\bar{\phi},\delta W_{i}]]\,,\label{eigenfrequency-W-full}\\
-\omega_{\phi}^{2}\delta\phi &=\bar{D}_{i}\bar{D}_{i}\delta\phi - e^{2}\,[\bar{\phi},[\bar{\phi},\delta\phi]] - \frac{4\lambda}{y}\,\text{Tr}\left(\bar{\phi}\delta\phi\right)\bar{\phi}\,.\label{eigenfrequency-phi-full}
\end{align}

In the sequence, we shall show that $\omega_{\phi}^{2}$ is always greater than zero, which means that the Higgs field perturbation does not contribute to any unstable mode. This assertion is also related to the fact that the linearized equations for perturbations of the gauge and Higgs fields decouple for asymptotic non-BPS monopole configurations.

Let us start by recalling the fact that we want the new (perturbed) configuration $\phi=\bar{\phi}+\delta\phi$, $W_{i}=\bar{W}_{i}+\delta W_{i}$ to have finite energy, which implies that it must satisfy $D_{i}\phi=0$ asymptotically. This means that, up to first order,
\begin{equation*}
D_{i}\phi=\bar{D}_{i}\left(\bar{\phi}+\delta\phi\right)+ie\,[\delta W_{i},\bar{\phi}]=0\,.
\end{equation*}
Moreover, recalling that the original (unperturbed) asymptotic configuration was such that $\bar{D}_{i}\bar{\phi}=0$, we get that 
\begin{equation*}
\bar{D}_{i}\delta\phi+ie\,[\delta W_{i},\bar{\phi}]=0\,.
\end{equation*}
Then, acting with $\bar{D}_{i}$ on the above equation and making use of \cref{background-gauge-condition} we obtain that
\begin{equation*}
\bar{D}_{i}\bar{D}_{i}\delta\phi-e^{2}\,[\bar{\phi},[\bar{\phi},\delta\phi]]=0\,.%\label{phi-simplification}
\end{equation*}
Using this result into \cref{eigenfrequency-phi-full}, we see that 
\begin{equation}
\omega_{\phi}^{2}\delta\phi=\frac{4\lambda}{y}\text{Tr}\left(\bar{\phi}\delta\phi\right)\bar{\phi}\,.\label{phi-frequency-simple}
\end{equation}
Thus, we only have two possibly interesting cases to analyze when $\delta\phi\neq0$: 
\begin{enumerate}
\item $\text{Tr}\left(\bar{\phi}\delta\phi\right)=0$, which implies that $\omega_{\phi}^{2}=0$. 
\item $\text{Tr}\left(\bar{\phi}\delta\phi\right)\neq0$. Then, we can multiply both sides of \cref{phi-frequency-simple} by $\frac{1}{y}\bar{\phi}$ and take the trace in order to see that 
\begin{equation}
\omega_{\phi}^{2}=4\,\lambda\frac{\text{Tr}\left(\bar{\phi}\bar{\phi}\right)}{y}=4\lambda\,v^{2}\geq0\,.
\end{equation}
\end{enumerate}
Therefore, the Higgs field perturbations cannot generate any unstable mode. Hence, from now on we shall take $\delta\phi=0$, in order to simplify the calculations of the eigenmodes related to the gauge field. Note that, after setting $\delta\phi=0$, we have a simpler background gauge condition derived from \cref{background-gauge-condition}. It reads,
\begin{equation}
\bar{D}_{i}\delta W_{i}=0\,.\label{short-backgroung-gauge}
\end{equation}

In order to analyze the spectrum of \cref{eigenfrequency-W-full}, we shall investigate its symmetries first. Let $L_{i}=-i\epsilon_{ijk}x^{j}\partial_{k}$ be the orbital angular momentum and $M_{i}$, $i=1,2,3$, the monopole generators. Since, 
\begin{align*}
[L_{i},L_{j}] & =i\epsilon_{ijk}L_{k}\,,\\{}
[M_{i},M_{j}] & =i\epsilon_{ijk}M_{k}\,,
\end{align*}
we can use the standard rules for addition of angular momentum to define 
\begin{equation*}
J_{i}\equiv L_{i}+M_{i}=-i\epsilon_{ijk}x^{j}\bar{D}_{k}+\frac{x^{i}}{r^{2}}\,x^{m}M_{m}\,,
\end{equation*}
where in the last equality we have used the asymptotic form of the original Dark Monopole gauge field given by \cref{ansatz-DM}. 

Further, we shall define the generalized angular momentum $\tilde{J}_{i}$ to be 
\begin{equation*}
\tilde{J}_{i}=J_{i}+S_{i}\,,
\end{equation*}
where $S_{i}$ is the spin operator, so that 
\begin{equation*}
\left(S_{i}\right)_{jk}\delta W_{k}=-i\epsilon_{ijk}\delta W_{k}\,,
\end{equation*}
which implies that $\left(S_{i}\right)_{jl}\left(S_{i}\right)_{lk}\,\delta W_{k}=2\,\delta W_{j}$.
Then, making use of \cref{short-backgroung-gauge} as well as the choice $x^{i}\delta W_{i}=0$ (associated with the fact that there is no radial gauge field component), one can see that 
\begin{equation}
\begin{aligned}
\left(\tilde{J}^{2}\right)_{ij}\delta W_{j}
&=-r^{2}\bar{D}_{m}^{2}\delta W_{i}+\partial_{r}\left(r^{2}\partial_{r}\right)\delta W_{i}+\frac{1}{r^{2}}\,[x^{l}M_{l},[x^{k}M_{k},\delta W_{i}]]\\
&+2\,ie\,\epsilon_{ijk}\,\frac{x^{j}}{r^{2}}[x^{l}M_{l},\delta W_{k}]\,. \label{D_mD_m}
\end{aligned}
\end{equation}
Then, the result of \cref{D_mD_m} can be used to rewrite \cref{eigenfrequency-W-full} as 
\begin{equation}
-\omega_{{\scaleto{G}{3.5pt}}}^{2}\,\delta W_{i}=\frac{1}{r^{2}}\,\partial_{r}\left(r^{2}\partial_{r}\right)\delta W_{i}-\frac{\left(\tilde{J}^{2}\right)_{ij}\delta W_{j}-[n^{l}M_{l},[n^{k}M_{k},\delta W_{i}]]}{r^{2}}-e^{2}\,[\bar{\phi},[\bar{\phi},\delta W_{i}]]\,.\label{eq-delta-W-with-mass}
\end{equation}
Moreover, for the perturbed solution to have finite energy, $D_{i}\phi=0$ for $r>R_{c}$. Now, since $\delta\phi=0$, we obtain that $[\delta W_{i},\bar{\phi}]=0$. This means that at each point in space $\delta W_{i}$ lies in the unbroken gauge group $G_{0}(\hat{r})$ and the mass term $[\bar{\phi},[\bar{\phi},\delta W_{a}]]$ in
\cref{eq-delta-W-with-mass} is zero. Hence, we only have to deal with 
\begin{equation}
-\omega_{{\scaleto{G}{3.5pt}}}^{2}\,\delta W_{i}=\frac{1}{r^{2}}\,\partial_{r}\left(r^{2}\partial_{r}\right)\delta W_{i}-\frac{\left(\tilde{J}^{2}\right)_{ij}\delta W_{j}-[n^{l}M_{l},[n^{k}M_{k},\delta W_{i}]]}{r^{2}}\,.\label{eq-delta-W-without-mass}
\end{equation}
This is the expected result since only long-range gauge perturbations can interfere in the asymptotic linear stability analysis.

Finally, note that we can use the gauge covariance of \cref{eq-delta-W-without-mass} to perform a gauge transformation associated with $g(\theta,\varphi)$, given by \cref{g-theta-phi}, in order to proceed with calculations in a more convenient way. This transformation is such that 
\begin{equation*}
n^{i}M_{i} = g(\theta,\varphi)\,M_{3}\,g^{-1}(\theta,\varphi)\,.
\end{equation*}
Then, we define
\begin{align*}
\mathcal{J}_{i} 
&= g^{-1}(\theta,\varphi)\,\tilde{J}_{i}\,g(\theta,\varphi)\,,\\
\delta \mathcal{W}_{i} 
&= g^{-1}(\theta,\varphi)\,\delta W_{i}\,g(\theta,\varphi)\,,
\end{align*}
so that we can write \cref{eq-delta-W-without-mass} as 
\begin{equation}
-\omega_{{\scaleto{G}{3.5pt}}}^{2}\,\delta \mathcal{W}_{i}=\frac{1}{r^{2}}\,\partial_{r}\left(r^{2}\partial_{r}\right)\delta \mathcal{W}_{i}-\frac{\left(\mathcal{J}^{2}\right)_{ij}\delta \mathcal{W}_{j}-[M_{3},[M_{3},\delta \mathcal{W}_{i}]]}{r^{2}}\,.\label{eq-delta-W-abelian-gauge}
\end{equation}
The convenience lies in the fact that the expressions for the perturbations $\delta\mathcal{W}_{a}$ are much simpler and lead directly to the solution of $\delta W_{i}$. 

\section{Reduction to Radial Equation}\label{Radial-Equations-Section}

A careful analysis shows that \cref{eq-delta-W-abelian-gauge} has a useful set of symmetries which reduce our problem to a radial equation since the differential operator acting on $\delta\mathcal{W}_{i}$ commutes with $\mathcal{J}_{3},\mathcal{J}^{2}$ and $M_{3}$.  Then, we can work with perturbations that are written in terms of simultaneous eigenfunctions of these generators.

Moreover, recalling the fact that $\delta\mathcal{W}_{i}$ belongs to $L(G_{0})$, we can write
\begin{equation}
\delta\mathcal{W}_{i} = \sum_{a}P_{a}(r)\mathcal{Y}_{ia}^{\,qJM}(\theta,\varphi)\,,\label{ansatz}
\end{equation}
where $P_{a}(r) \in \mathbb{R}$ is a radial function such that $P_{a}(R_{c}) = 0$ and $P_{a}(r\to\infty)=0$. Furthermore, we define
\begin{equation}
\mathcal{Y}_{ia}^{\,qJM} = Y_{i}^{qJM}T_{a}^{(q)}\,,\label{generalized-monopole-harmonics}
\end{equation} 
where $Y_{i}^{qJM}$ are the monopole vector spherical harmonics \cite{BrandtNeri,OlsenWu,Boulware,James,WeinbergHarmonics}
\begin{equation}
Y_{i}^{qJM}(\theta,\varphi) = \sum_{m=-j}^{+j}\sum_{m'=-1}^{+1}\,\braket{JM|j,m;1,m'}\,Y^{qjm}(\theta,\varphi)\,\chi_{i}^{m'}\,,\label{Monopole_Harmonics}
\end{equation}
with $Y^{qjm}(\theta,\varphi)$ being the monopole spherical harmonics of Wu and Yang \cite{WuYang} and 
\begin{equation}
\mathbf{\chi}^{-1} = \frac{1}{\sqrt{2}}
\begin{pmatrix}
1\\
-i\\
0
\end{pmatrix}\,,\quad
\mathbf{\chi}^{0} = \frac{1}{\sqrt{2}}
\begin{pmatrix}
0\\
0\\
1
\end{pmatrix}\,,\quad
\mathbf{\chi}^{1} = \frac{1}{\sqrt{2}}
\begin{pmatrix}
-1\\
-i\\
0
\end{pmatrix}\,,\label{chi}
\end{equation}
while $T_{a}^{(q)}$ are generators in $L(G_{0})$ such that
\begin{equation}
\left[M_{3},T_{a}^{(q)}\right] = q\,T_{a}^{(q)}\,.\label{charge-eigenvector}
\end{equation}
Then, the functions of \cref{generalized-monopole-harmonics} satisfy
\begin{subequations}
\begin{align}
\left(\mathcal{J}^{2}\right)_{ij}\mathcal{Y}_{j}^{qJM}
&=J\left(J+1\right)\mathcal{Y}_{i}^{qJM}\,,\label{j-eigenvalue}\\
\left(\mathcal{J}_{3}\right)_{ij} \mathcal{Y}_{j}^{qJM}
&=M\,\mathcal{Y}_{i}^{qJM}\,,\label{m-eigenvalue}\\
[M_{3},\mathcal{Y}_{i}^{qJM}]
&=q\,\mathcal{Y}_{i}^{qJM}\,,\label{q-eigenvalue}
\end{align}
\end{subequations}
where $J=|q|-1,|q|,|q|+1,\dots$, except when $|q|=0$ or $|q|=\frac{1}{2}$, where the minimum value of $J$ starts at $J_{min}=|q|$. As usual, $-J\leq M \leq J$, while we shall analyze the possible values of $|q|$ in \cref{Unstable-Modes-Section}. The range of values for $J$ comes directly from the standard rules for addition of angular momentum \cite{SWeinbergBook,Coleman50}. 

Then, after plugging \cref{ansatz} into \cref{eq-delta-W-abelian-gauge} and using the results \cref{j-eigenvalue,m-eigenvalue,q-eigenvalue}, we obtain that
\begin{equation}
-\omega_{{\scaleto{G}{3.5pt}}}^{2}\,P_{a}(r)=\frac{1}{r^{2}}\,\frac{d}{dr}\left(r^{2}\frac{d}{dr}\right)P_{a}(r)-\frac{J(J+1)-q^{2}}{r^{2}}\,P_{a}(r)\,,\label{original-ODE}
\end{equation} 
with the aforementioned boundary conditions. We know that a linear instability happens when $\omega_{{\scaleto{G}{3.5pt}}}^{2}<0$, i.e., when the time-dependence has the form of $\exp\left(|\omega_{{\scaleto{G}{3pt}}}|t\right)$. Now, we shall proceed with an analysis of the radial differential equation \eqref{original-ODE}. 

First, we make a change of variables from $r$ to the dimensionless
$\xi=evr$ and write the equation in the form of a Sturm-Liouville\footnote{Note, however, that this is not the ordinary Sturm-Liouville case since the interval is semi-infinite.} problem 
\begin{equation}
-\left[\xi^{2}\,P'_{a}(\xi)\right]'+[J(J+1)-q^{2}]\,P_{a}(\xi)=\bar{\omega}^{2}\,\xi^{2}\,P_{a}(\xi)\,,\label{sturm-liouville-form}
\end{equation}
where 
\begin{equation*}
\bar{\omega}^{2}\equiv\frac{\omega_{{\scaleto{G}{3.5pt}}}^{2}}{e^{2}v^{2}}\,.%\label{bar-omega}
\end{equation*}
Once more we recall that the boundary conditions are $P(\xi_{c})=0=P(\xi\to\infty)$, while \emph{primes} denote derivatives with respect to $\xi$. In the sequence, let us make $P_{a}(\xi)=\xi^{-1/2}f_{a}(\xi)$ so that \cref{sturm-liouville-form} gives
\begin{equation}
\xi^{2} f''_{a}(\xi) + \xi f'_{a}(\xi) + \left(\bar{\omega}^{2}\xi^{2}-\sigma^{2}\right)f_{a}(\xi)=0\,,\label{Bessel-General}
\end{equation}
with
\begin{equation*}
\sigma = \sqrt{\left(J+1/2\right)^{2}-|q|^{2}}\,.
\end{equation*}
Note that \cref{Bessel-General} is in the form of a Bessel equation \cite{AbramowitzStegun} and that $\sigma$ is either real or purely imaginary. Then, we can analyze the solutions for the possible values $J$ can assume.

First, when $J\geq|q|$, i.e., when $|q|=0\text{ or }1/2$, we see that $\sigma\in\mathbb{R}$ implying that the solution takes the form of
\begin{equation}
P_{a}(\xi) = C_{1}\xi^{-1/2}\left[J_{\sigma}(\bar{\omega}\xi)Y_{\sigma}(\bar{\omega}\xi_{c}) - J_{\sigma}(\bar{\omega}\xi_{c})Y_{\sigma}(\bar{\omega}\xi)\right]\,,\label{first-case}
\end{equation} 
where $C_{1}$ is an arbitrary constant, $\xi_{c}\leq\xi\leq\infty$ while $J_{\sigma}(\bar{\omega}\xi)$ and $Y_{\sigma}(\bar{\omega}\xi)$ denote the Bessel functions of the first and second kind, respectively\footnote{Note that, when $\sigma$ is not an integer or half-integer, the solution is written with $J_{-\sigma}(\bar{\omega}\xi)$ instead of $Y_{\sigma}(\bar{\omega}\xi)$}. Note that we have already used our first boundary condition. Now, one can verify that if $\bar{\omega}\in\mathbb{R}$ the spectrum is continuous and the condition at $\xi\to\infty$ does not impose any new constraint to the solution. On the other hand, if we look for a solution where $\bar{\omega}$ is purely imaginary, we find that \cref{first-case} can be written, up to a phase, in terms of the modified Bessel functions of real order, which implies that the boundary conditions cannot be satisfied simultaneously. This happens since a solution of the form of \cref{first-case}, written to satisfy the boundary condition at $\xi_{c}$, will diverge as $\xi\to\infty$. Therefore, there are no unstable modes in this situation.

In the sequence, we can analyze the case where $J=|q|-1$, associated with $|q|\geq1$. In this case, $\sigma=i\nu$, that is
\begin{equation*}
\nu = \sqrt{|q|-1/4}\,,\label{nu}
\end{equation*}
where now $\nu\in \mathbb{R}$. Using the results of \cite{Dunster}, one can show that this problem has a continuous spectrum when $\bar{\omega}^{2}>0$. Nevertheless, our interest here lies in the case of $\bar{\omega}^{2}<0$, so that \cref{Bessel-General} takes the form
\begin{equation}
\xi^{2} f''_{a}(\xi) + \xi f'_{a}(\xi) - \left(|\bar{\omega|}^{2}\xi^{2}-\nu^{2}\right)f_{a}(\xi)=0\,.\label{Unstable_Modes}
\end{equation}
This implies that solutions take the form of modified Bessel functions of purely imaginary order \cite{AbramowitzStegun}. We choose to write a general solution in terms of the linearly independent solutions \cite{Dunster}
\begin{align*}
L_{\mu}(z) &= \frac{i\pi}{2\sin(\mu\pi)}\,\left[I_{\mu}(z)+I_{-\mu}(z)\right]\,,\\
K_{\mu}(z) &= \frac{\pi}{2\sin(\mu\pi)}\,\left[I_{-\mu}(z)-I_{\mu}(z)\right]\,,
\end{align*} 
where $I_{\mu}(z)$ stands for the modified Bessel function of the first kind, for $\mu\neq 0$, $\mu\in\mathbb{C}$ and a complex variable $z$. Now, these functions possess the desirable property that, for the special case of our real variable $\xi$ with $\nu\in\mathbb{R}_{+}^{\ast}$, it follows that $L_{i\nu}(\xi), K_{i\nu}(\xi) \in \mathbb{R}$. This implies that a general solution to \cref{Unstable_Modes}, written in terms of $P_{a}(\xi)$, is given by
\begin{equation*}
P_{a}(\xi) = \xi^{-1/2}\left[A_{1a}L_{i\nu}(|\bar{\omega}|\xi) + A_{2a}K_{i\nu}(|\bar{\omega}|\xi)\right]\,,
\end{equation*}
with $A_{1a},A_{2a}\in\mathbb{R}$. However, from the asymptotic behavior of these functions as $\xi\to\infty$
\begin{align*}
L_{i\nu}(|\bar{\omega}|\xi)
&\approx \frac{1}{\sinh(\nu\pi)}\,\sqrt{\frac{\pi}{2|\bar{\omega}|\xi}}\,\exp(|\bar{\omega}|\xi)\left\{1 + \mathcal{O}\left(\frac{1}{|\bar{\omega}|\xi}\right)\right\}\,,\\
K_{i\nu}(|\bar{\omega}|\xi)
&\approx \sqrt{\frac{\pi}{2|\bar{\omega}|\xi}}\,\exp(-|\bar{\omega}|\xi)\left\{1 + \mathcal{O}\left(\frac{1}{|\bar{\omega}|\xi}\right)\right\}\,,
\end{align*}
we see that only $K_{i\nu}(|\bar{\omega}|\xi)$ is able to satisfy our condition at infinity.

Therefore, the solution to $P_{a}(\xi)$ is going to be of the form
\begin{equation}
P_{a}(\xi) = A_{1a}\,\xi^{-1/2}\,K_{i\nu}(|\bar{\omega}|\xi)\,.\label{General-Eingenfunction}
\end{equation}
The boundary condition $P_{a}(\xi_{c})=0$ implies that $|\bar{\omega}|\xi_{c}$ must be a root of $K_{i\nu}(|\bar{\omega}|\xi)$. From \cite{Dunster}, we know that any root $k_{n}$, $n=1,2,\dots$, is such that $0<k_{n}<\nu$. This means that $|\bar{\omega}|= k_{n}/\xi_{c}$, which leads to
\begin{equation*}
\bar{\omega}^{2} = -\left(\frac{k_{n}}{\xi_{c}}\right)^{2}\,.
\end{equation*}
That is, there will be one negative squared-eigenfrequency related with each one of the roots of $K_{i\nu}(|\bar{\omega}|\xi)$. In the case of our $SU(N)$ Dark Monopoles \cite{DeglmannKneipp}, $\xi_{c} = |\lambda_{p}|$ implying that
\begin{equation}
\bar{\omega}^{2} = -\left(\frac{k_{n}}{|\lambda_{p}|}\right)^{2}\,.\label{negative-eigenfrequency}
\end{equation}
Then the task of calculating the possible negative squared-eigenfrequencies comes down to the calculation of the roots of $K_{i\nu}(|\bar{\omega}|\xi)$. Moreover, up to an overall normalization, the eigenfunctions associated with these eigenmodes will be of the form of \eqref{General-Eingenfunction}.

\section{Dark Monopoles Unstable Modes}\label{Unstable-Modes-Section}

In order to find the Dark Monopole unstable modes explicitly, we shall first determine the generators $T_{a}^{(q)}$ associated with $|q|\geq 1$. With this aim, let us write
\begin{equation*}
M_{3} = 2T_{2}^{\alpha} = 2\,S\,T_{3}^{\alpha}\,S^{-1} = S\,\left(\alpha\cdot H\right)\, S^{-1}\,,
\end{equation*}
with $S = \exp\left(i\frac{\pi}{2}\,T_{1}^{\alpha}\right)\,\in L(G_{0})$ and $\alpha^{2}=2$, which is true for all roots of $su(n)$.
Then, for each $H_{i},\,E_{\beta}\in L(G_{0})$, we can define 
\begin{subequations}
\begin{align}
h_{i} &= S\,H_{i}\,S^{-1}\,,\label{little-H}\\
e_{\beta} &= S\,E_{\beta}\,S^{-1}\,,\label{little-E}
\end{align}
\end{subequations} 
where $h_{i},e_{\beta} \in L(G_{0})$. Note that, only here, $i$ ranges from $1$ to $r$, the rank of $L(G)$. Then, $M_{3}$ can be written as $M_{3} = \alpha\cdot h$, which implies that
\begin{equation}
\begin{aligned}
[M_{3},h_{j}] 
&= 0\,,\\
[M_{3},e_{\beta}] 
&= \left(\alpha\cdot\beta\right)e_{\beta}\,.\label{q-cartan}
\end{aligned}
\end{equation}
Comparing the result of \cref{q-cartan} with the one of \cref{charge-eigenvector}, we conclude that the unstable eigenmodes are the ones associated with generators $e_{\beta}$, where $\beta$ satisfy the condition
\begin{equation*}
|q| = |\alpha\cdot\beta|\geq 1\,,
\end{equation*}
Therefore, for a Dark Monopole associated with $M_{3}=2T_{2}^{\alpha}$, we can have unstable modes related to $e_{\pm\alpha}$ where $|q|=2$ and to $e_{\pm\beta}$ for a root $\beta$ such that $|\alpha\cdot\beta| = 1$, where $|q|=1$. Otherwise, $|q|=0$ and no instability is possible.

Now, recalling \cref{Monopole_Harmonics} together with the fact that
\begin{enumerate}[i)]
\item $\braket{JM|j,m;1,m'}^{\ast} = \braket{JM|j,m;1,m'}\,,$
\item $Y^{\ast\,qjm} = (-1)^{q+m}\,Y^{-q,j,-m}\,,$
\item $\chi_{i}^{m'\ast} = (-1)^{m'}\chi_{i}^{-m'}\,$
\end{enumerate}
one can show that, for $J=|q|-1$
\begin{equation}
Y_{i}^{\ast\,qJM}(\theta,\varphi) = (-1)^{q+M}\,Y_{i}^{-q,J,-M}(\theta,\varphi)\,. 
\end{equation}
And since we want $\delta\mathcal{W}^{\dagger}=\delta\mathcal{W}$, we find the solutions
\begin{equation}
\delta\mathcal{W}_{i}(\vec{r},t) = c\,\xi^{-1/2}\,K_{i\nu}\left(|\bar{\omega}|\xi\right)\,e^{+|\bar{\omega}|t}\left[Y^{qJM}_{i}e_{\beta}\pm Y^{-q,J,-M}_{i}e_{-\beta}\right]
\end{equation}
with the positive sign for the case when $q+M$ is even and the negative one for $q+M$ odd. Recalling \cref{nu} and the results above, $\nu$ is given by
\begin{equation}
\nu = 
\begin{cases}
\sqrt{3}/2,\text{ if }|q|=1\,,\\
\sqrt{7}/2,\text{ if }|q|=2\,.
\end{cases}\label{possible-nus}
\end{equation}
Furthermore, the negative-squared eigenfrequency $|\bar{\omega}|$ is given by \cref{negative-eigenfrequency}, depending directly on the roots of $K_{i\nu}(|\bar{\omega}|)$. These roots can be found numerically \cite{Mathematica}, where one can see that for $\nu = \sqrt{3}/2$ there is only one root at $|\bar{\omega}|\xi_{c} = 0.0329$, while $\nu=\sqrt{7}/2$ possess two roots, one at $|\bar{\omega}|\xi_{c} = 0.0145$ and another one at $|\bar{\omega}|\xi_{c} = 0.1565$. These values correspond to the possibilities of eigenfrequencies to our problem, while the eigenfunctions will start at one of these roots and go to zero at infinity. 

Finally, we show the explicit results for the case of a $SU(5)$ Dark Monopole, where $\xi_{c} = |\lambda_{p}|$ with $p=2$ or $3$ \cite{DeglmannKneipp}, so that $|\lambda_{p}| = \sqrt{6}/5$. The only eigenfrequency associated with $|q|=1$ is $\bar{\omega}_{|q|=1}^{2} = -0.0009$, while the two eigenfrequencies related to $|q|=2$ are $\bar{\omega}_{|q|=2}^{2} = -0.2041$ and $\bar{\omega}_{|q|=2}^{2} = -0.0018$.

\section{Discussions and conclusion}\label{Discussions-and-Conclusion-Section}

In this paper we have analyzed the asymptotic stability of the $SU(n)$ Dark Monopole solutions. Although these monopoles possess a conserved topological charge, it does not guarantee their stability.

We have showed that the unstable perturbations are associated with the generators of the unbroken group $G_{0}$ with eigenvalues $|q|=1$ and $|q|=2$ with respect to the monopole generator $M_{3}$. This result is analogous to the well-known criteria from the work of Brandt and Neri \cite{BrandtNeri} for non-abelian singular monopoles. We have obtained the explicit form of these unstable perturbations and the corresponding eigenfrequencies. 

In principle, the existence of a conservation law associated with an asymptotic symmetry of the field configuration could provide a dynamical stability to our solution. However, we saw that this is not the case. Thus, we expect this conserved non-abelian magnetic charge (in the $\zeta$-direction), to radiate away from the monopole. As a comparison, in the 't Hooft-Polyakov monopole the Killing vector is the Higgs field itself, implying it cannot be unwound by means of a continuous deformation without taking the asymptotic scalar field out of the vacuum manifold since this would require an infinite energy.

The fact that these $SU(n)$ Dark Monopoles are static and unstable indicates that they are saddle point solutions to the classical field equations, i.e., they are a stationary point of the energy, but not a minimum. %Since they lie in the trivial topological sector $\mathcal{C}_{0}$, where the field configuration with the minimal energy is the trivial vacuum, this result was somewhat expected. This also agrees with the non-existence of a BPS bound for these monopoles, which we have discussed at \cite{DeglmannKneipp}. Moreover, 
Saddle point solutions occur in many field theories (see, for instance, \cite{MantonSutcliffe,MantonRoyal}), where we take as examples the electroweak sphaleron of Manton and Klinkhamer \cite{MantonKlinkhamer} and the monopole-antimonopole pair of Taubes \cite{Taubes1,Taubes2}. 
 
Therefore, it would be interesting to investigate if the existence of Dark Monopoles can also be associated with a non-trivial topology of the configuration space, in a similar way to the work of James \cite{James2} for the electroweak strings.

\section*{Acknowledgments}

M.L.D. thanks Victor Espinoza for useful discussions. This study was financed in part by the Coordenação de Aperfeiçoamento de Pessoal de Nível Superior - Brasil (CAPES) - Finance Code 001.

\end{document}